\documentclass[conference]{IEEEtran}
\IEEEoverridecommandlockouts
\usepackage{amsmath,amssymb,amsfonts}
\usepackage[noadjust]{cite}
\usepackage{graphicx}
\usepackage{textcomp}
\usepackage{xcolor}
\usepackage{soul,color}          % highlight
\usepackage{mathrsfs}      % to use \mathscr
\usepackage{algorithm}
\usepackage[noend]{algpseudocode}
\usepackage{bm}
\usepackage{multirow}
\usepackage{slashbox}   % \backslashbox
\usepackage{makecell}   % break a line within a cell
\usepackage{varwidth}   % line break for loop

\usepackage{makecell}

\usepackage[switch]{lineno}
\usepackage{enumitem}
\setlist[itemize]{align=parleft,left=0pt..1em}

\usepackage{comment}

\def\BibTeX{{\rm B\kern-.05em{\sc i\kern-.025em b}\kern-.08em
    T\kern-.1667em\lower.7ex\hbox{E}\kern-.125emX}}

\setstcolor{red}

\setstcolor{red}

\renewcommand{\algorithmiccomment}[1]{\bgroup\hfill\tiny//~#1\egroup}
\algblockdefx[ForEach]{ForEach}{EndForEach}[1]{\textbf{for each} #1 \textbf{do}}%{\textbf{end for each}}

\begin{document}

\setlength{\intextsep}{0.75\baselineskip} % Adjust the space between the title and the author block

% \title{Traffic Engineering in Multi-band Optical Networks}
\title{Increasing Information-Carrying Capacity by Exploiting Diverse Traffic Characteristics in Multi-Band Optical Networks}

% \title{Increasing Information-Carrying Capacity by Exploiting Application Characteristics in Multi-Band Optical Networks}

% \begin{comment}
% For ANTS 2024
\author{
\hspace{-0.5cm}\IEEEauthorblockN{Ramanuja Kalkunte}
\IEEEauthorblockA{
% \hspace{-0.5cm}\footnotesize\textit{Dept. of Computer Science} \\
\footnotesize\hspace{-0.5cm}\textit{University of California,}\\
\hspace{-0.5cm}\textit{Davis}, USA \\
\hspace{-0.5cm}rkalkunte@ucdavis.edu}
\and
\hspace{-0.02cm}\IEEEauthorblockN{Forough Shirin Abkenar}
\IEEEauthorblockA{
% \hspace{0.02cm}\footnotesize\textit{Dept. of Computer Science} \\
\footnotesize\hspace{-0.02cm}\textit{University of California,}\\
\hspace{-0.02cm}\textit{Davis}, USA \\
\hspace{-0.2cm}fshirina@ucdavis.edu}

\and
\IEEEauthorblockN{Sifat Ferdousi}
\IEEEauthorblockA{
% \footnotesize\textit{Dept. of Computer Science} \\
\footnotesize\textit{University of California,}\\
\textit{Davis}, USA \\
sferdousi@ucdavis.edu}

\and
\hspace{-0.1cm}\IEEEauthorblockN{Rana Kumar Jana}
\IEEEauthorblockA{
% \hspace{0cm}\footnotesize\textit{Dept. of Electronics and Comm. Engg.} \\
\hspace{0cm}\footnotesize\textit{Indraprastha Institute of Information}\\
\hspace{0cm} \textit{Technology, Delhi}, India\\
\hspace{0cm}ranaj@iiitd.ac.in}

% \and
% \hspace{-0.05cm}\IEEEauthorblockN{Andrew Lord}
% \IEEEauthorblockA{\footnotesize\textit{British Telecom} \\
% Ipswich, UK \\
% \hspace{-0.05cm}andrew.lord@bt.com}

\and
\IEEEauthorblockN{Anand Srivastava}
\IEEEauthorblockA{
% \hspace{-0.8cm}\footnotesize\textit{Dept. of Electronics and Comm. Engg.} \\
\footnotesize\textit{Netaji Subhas University of Technology}\\
\textit{Delhi}, India\\
anand@nsut.ac.in}

\and
\IEEEauthorblockN{Abhijit Mitra}
\IEEEauthorblockA{
% \hspace{-0.2cm}\footnotesize\textit{Dept. of Electronics and Comm. Engg.} \\
\footnotesize\textit{Indraprastha Institute of Information}\\
\textit{Technology, Delhi}, India\\
abhijit@iiitd.ac.in}

\and
\IEEEauthorblockN{Massimo Tornatore}
\IEEEauthorblockA{
% \hspace{-0.5cm}\footnotesize\textit{Dept. of Electronics, Information,} \\
\footnotesize\textit{Politecnico di Milano}, Italy\\
% \hspace{-0.8cm}Italy \\
massimo.tornatore@polimi.it}

\and
\IEEEauthorblockN{Biswanath Mukherjee}
\IEEEauthorblockA{
% \footnotesize\textit{Dept. of Computer Science} \\
\footnotesize\textit{University of California,}\\
\textit{Davis}, USA \\
bmukherjee@ucdavis.edu}
}

% \end{comment}

\begin{comment}
\DeclareRobustCommand{\IEEEauthorrefmark}[1]{\smash{\textsuperscript{\footnotesize #1}}}

\author{\IEEEauthorblockN{\textbf{Ramanuja Kalkunte}\IEEEauthorrefmark{1}, \textbf{Forough Shirin Abkenar}\IEEEauthorrefmark{1},
\textbf{Rana Kumar Jana}\IEEEauthorrefmark{2},
\textbf{Davide Aureli}\IEEEauthorrefmark{3}, 
\textbf{Sifat Ferdousi}\IEEEauthorrefmark{1},\\
\textbf{Anand Srivastava}\IEEEauthorrefmark{2},
\textbf{Abhijit Mitra}\IEEEauthorrefmark{2},
\textbf{Massimo Tornatore}\IEEEauthorrefmark{4}, and
\textbf{Biswanath Mukherjee}\IEEEauthorrefmark{1}}
\IEEEauthorblockA{\small\IEEEauthorrefmark{1}University of California, Davis, USA \IEEEauthorrefmark{2}Indraprastha Institute of Information Technology, Delhi, India \IEEEauthorrefmark{3}INPS, Rome, Italy\\\IEEEauthorrefmark{4}Politecnico di Milano, Italy}
%\IEEEauthorblockA{}
\IEEEauthorblockA{\small Email: \{rkalkunte, fshirina, sferdousi, bmukherjee\}@ucdavis.edu, \{ranaj, anand, abhijit\}@iiitd.ac.in, davide.aureli@inps.it,\\massimo.tornatore@polimi.it}}
\end{comment}

\maketitle

\begin{abstract}

Efficient network management in optical backbone networks is crucial for handling continuous traffic growth. In this work, we address the challenges of managing dynamic traffic in C- and C+L-band optical backbone networks while exploring application flexibility, namely the compressibility and delayability metrics. We propose a strategy, named Delay-Aware and Compression-Aware (DACA) provisioning algorithm, which reduces blocking probability, thereby increasing information-carrying capacity of the network compared to baseline strategies.

\end{abstract}


\vspace{-1mm}
\section{Introduction}\label{sec:intro}
% \vspace{-0.22cm}

The growth of heterogeneous and bandwidth-hungry 5G/6G applications \cite{Cisco} calls for new efficient on-demand bandwidth-provisioning strategies in backbone networks. To support these diverse applications, optical backbone networks, which typically support quasi-static traffic, may soon need to evolve to adapt to increasingly dynamic traffic patterns \cite{ONDM_1}. The emergence of Elastic Optical Networks (EONs) has allowed operators to maximize utilization of the available C-band spectrum in Single-Mode Fibers (SMFs) \cite{JOCN_5}. However, to overcome the limited capacity of C band, optical backbone networks are migrating toward Multi-band (MB) transmission \cite{JLT_4}, starting with the L band due to matured technology. As such, operators are already expanding their infrastructure to support MB transmission for ever-growing service requirements \cite{Entel}.

Examining the characteristics of novel dynamic network services, two key flexibilities can be observed, \textit{Compressibility} and \textit{Delayability}. Delayability means that some services can be deferred in time, while compressibility means that the amount of bandwidth required by a connection can be reduced while still offering an acceptable level of service. Both delayability and compressibility can be leveraged by service providers and operators to accommodate more traffic at a given time. While some prior research has studied delayability \cite{ICTON_1}, others have focused on compressibility metrics \cite{TNET_1}. In this work, we differentiate traffic based on their application flexibility (see Section \ref{sec:traffic_types}) and highlight the \textit{joint impact} of both \textit{delayability and compressibility}. 
We propose a novel Delay-Aware and Compression-Aware (DACA) provisioning algorithm, which exploits delayability and compressibility characteristics of dynamic traffic to enhance the information-carrying capacity of the network. This enhancement is demonstrated by showing a reduction in Blocking Probability (BP).

% \vspace{-0.06cm}
\section{System Model}
\subsection{Diverse Traffic Types}
\label{sec:traffic_types}
\vspace{-1mm}

In this work, we categorize traffic into three major types based on its  delayability and compressibility metrics:

\begin{itemize}
    \item \textit{Type 1}: Non-Delayable and Non-Compressible, e.g., Google search. Statistics show that there are about 99,000 global search queries every second \cite{google_search}.
    \item \textit{Type 2}: Compressible,  e.g., video streaming.\\
    - \textit{2a}: Compressible and Delayable, e.g., on-demand videos provided by YouTube, Netflix, etc. Statistics reveal that 40\% of YouTube users in the UK access the platform daily \cite{YouTube_UK}.\\
    - \textit{2b}: Compressible and Non-Delayable, e.g., live stream on platforms such as Zoom and Microsoft Teams. 
    % Statistics show that about 17\% of Internet users watch live streaming content on a weekly basis in the UK \cite{live_stream_1}.
    Statistics show that about 27\% of Internet users watch live streaming content on a weekly basis \cite{live_stream}.

    % \item \textit{Type 3}: \rkr{Non-Compressible and Delayable}, e.g., data backups carried out by users and enterprises (i.e., full, incremental, etc). Due to variations in data backup characteristics (holding time and data rate), type 3 is divided into two sub-categories: \textit{3a (user backup)}, such as Google Drive and iCloud, and \textit{3b (enterprise backup)}, such as AWS and Microsoft Azure. On average, medium-sized companies benefit from backing up data every 24 hours \cite{file_backup}.
    
    % \item \textit{Type 3}: Non-Compressible and Delayable, e.g., data backups carried out by users and enterprises (i.e., full, incremental, etc). \\
    % \textit{3a}: User backup, such as Google Drive and iCloud. \\
    % \textit{3b:} Enterprise backup, such as AWS and Microsoft Azure. On average, medium-sized companies benefit from backing up data every 24 hours \cite{file_backup}.

    \item \textit{Type 3}: Delayable and Non-Compressible, e.g., data backup.\\
    - \textit{3a}: User backup, e.g., Google Drive and iCloud. \\
    - \textit{3b}: Enterprise backup, e.g., AWS and Microsoft Azure. On average, medium-sized companies benefit from backing up data every 24 hours \cite{file_backup}.

\end{itemize}

\vspace{-0.1mm}
\subsection{Network Model}\label{sec:network_model}
\vspace{-1mm}
We consider an elastic optical backbone network topology, $G(V,E)$, comprising $|V|$ nodes and $|E|$ links, where $V$ and $E$ represent set of nodes and set of links, respectively. Our study evaluates two scenarios: network operating in only C band and in C+L bands. Each band is composed of 133 channels with a frequency spacing of 37.5 GHz \cite{PNC_1}. 
%Considering various traffic types discussed in Section \ref{sec:traffic_types}, 
% \begin{table}[!t]
%     \centering
%     \caption{Traffic Parameters: (Type ($q$), Holding Time ($\tau$), Arrival rate ($\lambda$), Compression Factor ($\phi$), Delayability ($\delta$), and Data rate ($\gamma$)}
%      % \footnotesize
%      % \fontsize{14pt}{14pt}\selectfont
%      \LARGE
%      \resizebox{\columnwidth}{!}{
%     \begin{tabular}{|c|c|c|c|c|c|c|c|}
%         \hline
%         \thead{\bm{$q$}} & \thead{\textbf{Percentage of}\\\textbf{Total Requests}} & \thead{\bm{$\tau$}\\\textbf{(min)}} & \thead{\bm{$\lambda_{peak}$}\\\textbf{(requests/s)}} & \thead{\bm{$\lambda_{off-peak}$}\\\textbf{(requests/s)}} & \bm{$\phi$} & \thead{\bm{$\delta$}\\\textbf{(min)}} & \thead{\bm{$\gamma$}\\\textbf{(Gbps)}} \\
%         \hline
%         1 & 20\% & 5 & 8 & 2 & N/A & N/A & $\mathcal{U}$[100, 200] \\
%         \hline
%         2a & 45\% & $\mathcal{U}$[30, 90] & 100 & 25 & 50\% & $\mathcal{U}[3,5]$ & $\mathcal{U}$[200, 400]  \\
%         \hline
%         2b & 25\% & $\mathcal{U}$[20, 40] & 48 & 12 & 50\% & N/A & $\mathcal{U}$[200, 400]  \\
%         \hline
%         3a & 6\% & $\mathcal{U}$[8, 12] & 8 & 2 & N/A & $\mathcal{U}[2,4]$ & $\mathcal{U}$[100, 200]  \\
%         \hline
%         3b & 4\% & $\mathcal{U}$[360, 600] & 4 & 1 & N/A & $\mathcal{U}[360,720]$ & 400 \\
%         \hline
%     \end{tabular}}
%     \label{tab:traffic_parameters}
% \end{table}
We consider that the incoming traffic  is dynamic in nature and remains in the network for a given period of time.
Considering various traffic types discussed in Section \ref{sec:traffic_types}, 
there is a total of $R$ requests, where each request $r\in R$ is described using a tuple $(s, d, \gamma, \overline{\gamma}, q, \phi, \delta, \tau)$. Here, $s$ and $d$ denote source and destination nodes, respectively, and they are generated using a gravity model consisting of traffic generation probability of each node (see Fig. \ref{fig:BT_UK}).
Parameter $\gamma$ represents the required data rate and $\overline{\gamma}$ is the minimum acceptable data rate defined by the Service Level Agreement (SLA) between the users and the service providers. 
Also, $q$ represents the traffic type (refer to Section \ref{sec:traffic_types}), $\phi$ is the scaling factor to compress the data rate ($\gamma$), $\delta$ is the maximum tolerable delay, and $\tau$ indicates the holding time of the request. Note that parameters are adjusted for different traffic types; e.g., as shown in Table \ref{tab:traffic_parameters}, type 2b traffic has $\phi = 50\%$ and $\delta$ marked as Not Applicable (N/A), indicating that it is compressible up to 50\% but non-delayable.

In this study, each request is assumed to occupy one lightpath; and, to provision these lightpaths, their SLA must be satisfied. SLA compliance is ensured by providing adequate Quality of Transmission (QoT), which is often quantified in optical networks using Generalized Signal-to-Noise Ratio (GSNR). The data rate and modulation format of each lightpath is obtained using a GSNR window, as specified in \cite{PNC_1}. To calculate GSNR based on the current network state, we employ two Machine Learning (ML)-based QoT estimators: one for C-band-only and one for C+L bands. The physical-layer model and the ML models for QoT estimation are adopted from \cite{PNC_1}. 
%Note that dynamic traffic can have greater impact on the GSNR of lightpaths, which may lead to SLA violations and consequently cause connections to be dropped. This study, however, focuses solely on the number of blocked connections (BP).
Note that dynamic traffic can cause frequent changes in the network state resulting in variations in the GSNR of the lightpaths. This may lead to SLA violations causing some connections to be dropped. This study, however, focuses solely on the number of blocked connections (BP).

\vspace{-0.02in}
\section{Delay- and Compression-Aware Provisioning}
%Delay and Compression based Strategy
\label{sec:proposed_strategies}

Delayability and Compressibility have significant impact on traffic provisioning. We propose an algorithm, namely Delay-Aware and Compression-Aware (DACA), to jointly exploit these characteristics with the objective to reduce BP, and hence increase the information-carrying capacity of the network. This strategy utilizes the delayability and compressibility of certain incoming traffic to postpone their provisioning time and/or compress their bandwidth to accommodate them efficiently when spectral resources, i.e., Frequency Slots (FSs), are not available. Initially, DACA checks if it can provision a request $r$ in its uncompressed form; if the request cannot be provisioned due to unavailable slots, it is delayed. When the maximum tolerable delay ($\delta$) of $r$ has been exhausted and no slots are available, DACA compresses the request. Compression lowers the required data rate $\gamma$ by a factor of $\phi$, thereby reducing the number of FSs required to provision the request. If compression does not help, the request is blocked.

Algorithm \ref{alg:DACA} outlines DACA, which takes network topology ($G$) and a set of requests ($R$) as input. 
The number of blocked requests ($N_{b}$) is initialized to zero. We assume that, for a 24-hour period, peak hours of operation is between  $p_{s}$ and $p_{e}$, and the rest of the period is considered off-peak. GSNR of all active lightpaths is re-estimated every $t^{p}$ time unit during peak hours and every $t^{o}$ time unit during off-peak hours. After re-estimation, lightpaths for which the data rates ($\gamma$) are within their SLA ($ \overline{\gamma}$) are retained in the network. Routing and Spectrum Allocation (RSA) of the requests employ k-Shortest Path and First-Fit (FF) strategies, respectively. For each time unit $t$ and each incoming traffic $r \in R$, RSA identifies the candidate path(s) and the FSs. Each request is then checked and decisions are made based on the data rate ($\Gamma_{FS}$) of the available slot(s). If $\Gamma_{FS}$ is greater than the required data rate ($\gamma$), the path and FSs are assigned to the request. If not, request $r$, based on its traffic type $q$, is either delayed by 1 time unit or compressed to $\phi*\gamma$. Otherwise, $r$ is blocked. Once a request completes its service, i.e., its holding time ($\tau$), the lightpath is released. All type 3b traffic, for which [arrival time + delayability factor ($\delta$)] falls within $(p_{s},{p\prime}_{e})$, where  ${p\prime}_{e} = p_{e} + m$, will be postponed until ${p\prime}_{e}$ with $\delta = 0$.

\begin{table*}[htbp]
    \centering
    \caption{Traffic Parameters: Type ($q$), Holding Time ($\tau$), Arrival rate ($\lambda$), Compression Factor ($\phi$), Delayability ($\delta$), and Data rate ($\gamma$)}
     \footnotesize
    \begin{tabular}{|c|c|c|c|c|c|c|c|}
        \hline
        \bm{$q$} & \textbf{Percentage of Total Requests} & \bm{$\tau$} \textbf{(min)} & \bm{$\lambda_{peak}$} \textbf{(requests/s)} & \bm{$\lambda_{off-peak}$} \textbf{(requests/s)} & \bm{$\phi$} & \bm{$\delta$} \textbf{(min)} & \bm{$\gamma$} \textbf{(Gbps)} \\
        \hline
        1 & 20\% & 5 & 8 & 2 & N/A & N/A & $\{100, 200\}$ \\
        \hline
        2a & 45\% & $\mathcal{U}[30, 90]$ & 100 & 25 & 50\% & $\mathcal{U}[3,5]$ & $\{200, 400\}$  \\
        \hline
        2b & 25\% & $\mathcal{U}[20, 40]$ & 48 & 12 & 50\% & N/A & $\{200, 400\}$  \\
        \hline
        3a & 6\% & $\mathcal{U}$[8, 12] & 8 & 2 & N/A & $\mathcal{U}[2,4]$ & $\{100, 200\}$  \\
        \hline
        3b & 4\% & $\mathcal{U}[360, 600]$ & 4 & 1 & N/A & $\mathcal{U}[360,720]$ & 400 \\
        \hline
    \end{tabular}
    \label{tab:traffic_parameters}
    \vspace{-14pt}
\end{table*}
% \vspace{-1mm}
% ANTS 2024
\begin{algorithm}[H]
\caption{}\label{alg:DACA}
\small
\begin{algorithmic}[1]

\Statex \textbf{Input:} $G(V,E)$, $R$;
\Statex \textbf{Output:} $N_{b}$;
\State \textbf{Initialize:}  $N_{b} = 0$, $p_{s}$, $p_{e}$, ${p\prime}_{e}$, $t^{p}$, $t^{o}$; 
% , $P = \emptyset$;
\For{\text{each time $t$}}
\If {$p_{s} < t < p_{e}$}, re-estimate GSNR every $t^{p}$ time unit
\Else 
\State Re-estimate GSNR every $t^{o}$ time unit
\EndIf
\For{\text{incoming request $r \in R$}}
% \State $provision\_flag$ = \textbf{False}
\State
\begin{varwidth}[t]{\linewidth}
\par Perform corresponding RSA;
\end{varwidth}\label{line_provision_start}

\If {\text{slot(s) available} and {$\Gamma_{FS} \geq \gamma$}}
\State Assign slot(s) to request $r$;\label{line_provision_end}
% \State Assign path $p$ to request $i$;
% $P \leftarrow p$;
\Else
% \If{$q == 1$}
% Block $i$; $N_{b}~+= 1$;
% \Else
\If{$q == 2a$}
\If{$\delta > 0$}
    delay provisioning; $\delta ~-= 1$;
\Else~~$\gamma = \phi*\gamma$; perform lines \ref{line_provision_start}-\ref{line_provision_end};
%     \State Perform corresponding RSA;

%     \If {\text{no slot available}}
% Block $i$; $N_{b}~+= 1$;
\EndIf % \delta > 0
% \EndIf
\ElsIf{$q == 2b$}
\State $\gamma = \phi*\gamma$; perform lines \ref{line_provision_start}-\ref{line_provision_end};
%     \State Perform corresponding RSA;
%     \If {\text{no slot available}}
% Block $i$; $N_{b}~+= 1$;
    
% \EndIf

\ElsIf{$q == 3a$}
\If{$\delta > 0$}
    delay provisioning; $\delta ~-= 1$;
\Else~~perform lines \ref{line_provision_start}-\ref{line_provision_end};
% \Else~~Block $i$; $N_{b}~+= 1$;
\EndIf % \delta > 0

\ElsIf{$q == 3b$}
\If{$p_{s} < (t + \delta) < {p\prime}_{e}$}
delay $r$ till ${p\prime}_{e}$; $\delta = 0$;
\ElsIf{$\delta > 0$}
    delay provisioning; $\delta ~-= 1$;
\Else~~perform lines \ref{line_provision_start}-\ref{line_provision_end};
% \Else~~Block $i$; $N_{b}~+= 1$;
\EndIf % t < 22:00
\EndIf % q == 2a
% \EndIf % else q != 1

% \If{$provision\_flag$}
% \State Perform corresponding RSA;
% \If {\text{no slot available}}
% Block $i$; $N_{b}~+= 1$;
% \EndIf % provision_flag
\State Block $r$; $N_{b}~+= 1$;
\EndIf % else (slots are not available)
\EndFor % request i
% \If {$t_{s}^{p} \leq t \leq t_{e}^{p}$}, re-estimate GSNR every $t_{p}$ time unit
% \Else 
% \State Re-estimate GSNR every $t_{o}$ time unit
% \EndIf
\EndFor % time t
\State\textbf{Return} $N_{b}$
\end{algorithmic}
\end{algorithm}

\vspace{-0.18in}
\section{Numerical Evaluation}\label{sec:results}
\vspace{-0.02in}
\subsection{Modeling and Simulation Setup}
% \vspace{-0.05cm}
% An event-driven, custom-built Python simulator is used to model (i.e., emulate) a dynamic traffic environment in a C-/C+L-band framework, while incorporating physical-layer model from \cite{PNC_1}. 
% We use, as reference topology, the BT-UK network (see Fig. \ref{fig:BT_UK}), which comprises 22 nodes and 35 bi-directional links, where the average link length is about 147 km \cite{JOCN_5}. We repeat and average the simulations for 10 seeds, each with 15,000 demands. Network traffic follows a Poisson distribution with arrival rates of  $\lambda_{peak}$ and $\lambda_{off-peak}$, which varies across each traffic type, during peak and off-peak hours, respectively.
% Table \ref{tab:traffic_parameters} presents all traffic parameters, where $\mathcal{U}[\cdot,\cdot]$ denotes a uniform distribution. 
% Traffic is generated over multiple days. Figure \ref{fig:Utilization_C_C+L} shows a 24-hour snapshot with peak hours from $p_{s} = 8\text{am}$ to $p_{e} = 8\text{pm}$. For traffic type 3b, $m = 2$, such that ${p\prime}_{e} = 10\text{pm}$. We set $t^{p} = 10$ and $t^{o} = 100$.

\begin{figure}[!t]
    \centering
    \includegraphics[width=0.9\columnwidth]{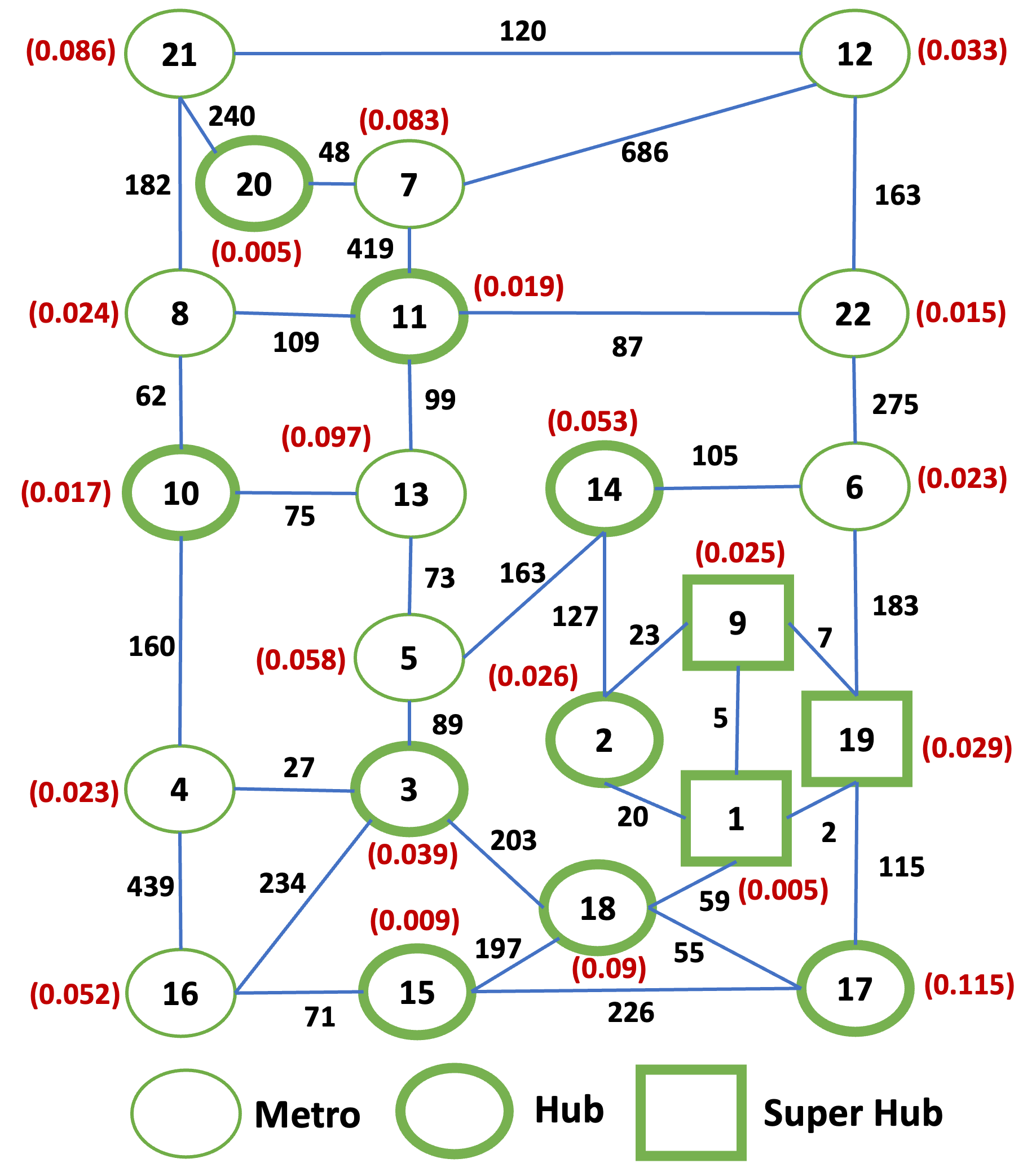} 
    % \vspace{-6pt}
    \caption{BT-UK network: link lengths in kilometer (km) and traffic generation probabilities in parentheses \cite{PNC_1, JOCN_4}.}
    \label{fig:BT_UK}
    \vspace{-15pt}
\end{figure}

An event-driven, custom-built Python simulator is used to model (i.e., emulate) a dynamic traffic environment in a C-/C+L-band framework, while incorporating physical-layer model from \cite{PNC_1}. 
We use, as reference topology, the BT-UK network (see Fig. \ref{fig:BT_UK}), which comprises 22 nodes and 35 bi-directional links, where the average link length is about 147 km \cite{JOCN_5}. We repeat and average the simulations for 10 seeds, each with 15,000 demands. Network traffic follows a Poisson distribution with arrival rates of  $\lambda_{peak}$ and $\lambda_{off-peak}$, which varies across each traffic type, during peak and off-peak hours, respectively.
Table \ref{tab:traffic_parameters} presents all traffic parameters, where $\mathcal{U}[\cdot,\cdot]$ denotes a uniform distribution.
% \rkr{v1: For $\gamma$, $\{\cdot\}$ indicates the set of candidate values, among which a value is selected uniformly.}\\
% \rkr{v2: For $\gamma$, the set $\{x, y\}$ indicates that $x$ or $y$ is selected with equal probability.}\\
For $\gamma$, $\{\cdot\}$ indicates that an element in a set is selected uniformly.
Traffic is generated over multiple days. Figure \ref{fig:Utilization_C_C+L} shows a 24-hour snapshot with peak hours from $p_{s} = 8\text{am}$ to $p_{e} = 8\text{pm}$. For traffic type 3b, $m = 2$, such that ${p\prime}_{e} = 10\text{pm}$. We set $t^{p} = 10$ and $t^{o} = 100$.

% $t_{s}^{3b} = 8~\text{a.m.}$, and $t_{e}^{3b} = 10~\text{p.m}$.}

% \vspace{-1mm}
\subsection{Baseline Strategies }
\vspace{-1mm}
To demonstrate the efficiency of DACA (see Section \ref{sec:proposed_strategies}), we compare it with a set of baseline strategies described below:

\begin{itemize}
    \item \noindent\textbf{No Delay, No Compression (NDNC)} 
    % This strategy processes all traffic uniformly such that no traffic is delayed or compressed. If incoming traffic cannot be provisioned, it is blocked.
    processes all traffic uniformly with no delay or compression.
    If incoming traffic cannot be provisioned, it is blocked.
    %unprovisioned traffic is blocked.
    % , thus maintaining consistent handling for every data stream.

    \item \noindent\textbf{Delay-Aware (DA)} only delays the provisioning of requests when resources are unavailable. If incoming traffic cannot be immediately provisioned and is delayable, DA postpones its provisioning. Note that provisioning of traffic type 3b follows the approach mentioned in DACA. 
    
    \item \noindent\textbf{Compression-Aware (CA)} only compresses the traffic when resources are unavailable. If incoming traffic cannot be provisioned and is compressible, CA implements compression and provisions the request in its compressed form. 
        
\end{itemize}

\vspace{-1mm}
\subsection{Performance Evaluation in C vs. C+L}
\vspace{-1mm}
\label{sec:eval_proposed_strategies}

\begin{figure}[!h]
    \centering
    \includegraphics[width=0.9\columnwidth]
    % {images/C_vs_C+L_3_300_dpi.png}
    {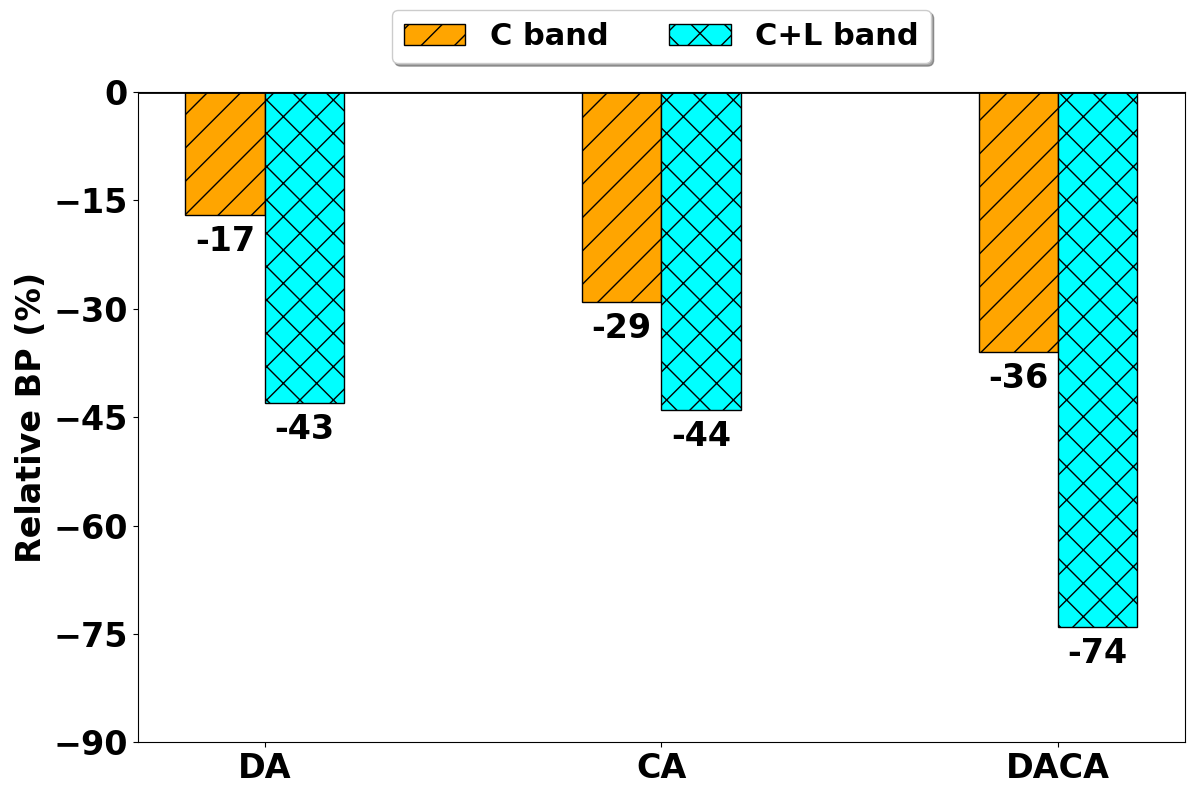}
    \vspace{-10pt}
    \caption{Relative BP induced by all strategies w.r.t.  NDNC in C and C+L.}
    \label{fig:C_C+L}
    \vspace{-2mm}
\end{figure}

We now evaluate the performance of DA, CA, and DACA w.r.t.  NDNC, as shown in Fig. \ref{fig:C_C+L}. For context, NDNC resulted in BP of 13.4\% of total network traffic for the C-band-only scenario. Relative to this, DA and CA lower the BP by 17\% and 29\%, respectively. This can be attributed to the fact that, while NDNC accommodates incoming requests by only checking for available bandwidth, DA and CA employ delayability and compressibility, respectively, which allows the network to operate with greater flexibility. However, with DACA, we observe a 36\% reduction in BP. In DACA, if a request cannot be provisioned after delaying, it may still be possible to provision in its compressed form. This allows the network to serve more connection requests compared to NDNC, DA, and CA. Similarly, activation of L band led to reduced BP of 4.6\% of total traffic in NDNC. Due to additional bandwidth in L band, note that DA, CA, and DACA show a significant improvement of 43\%, 44\%, and 74\%, respectively, in BP. The rest of this study will focus on the performance of these strategies in the C+L-band framework.

\vspace{-1mm}
\subsection{Performance Evaluation based on Traffic Types}
\begin{figure}[!t]
    \centering
    \includegraphics[width=\columnwidth]
    % {images/Relative_BP_NDNC_2_300_dpi.png}
    {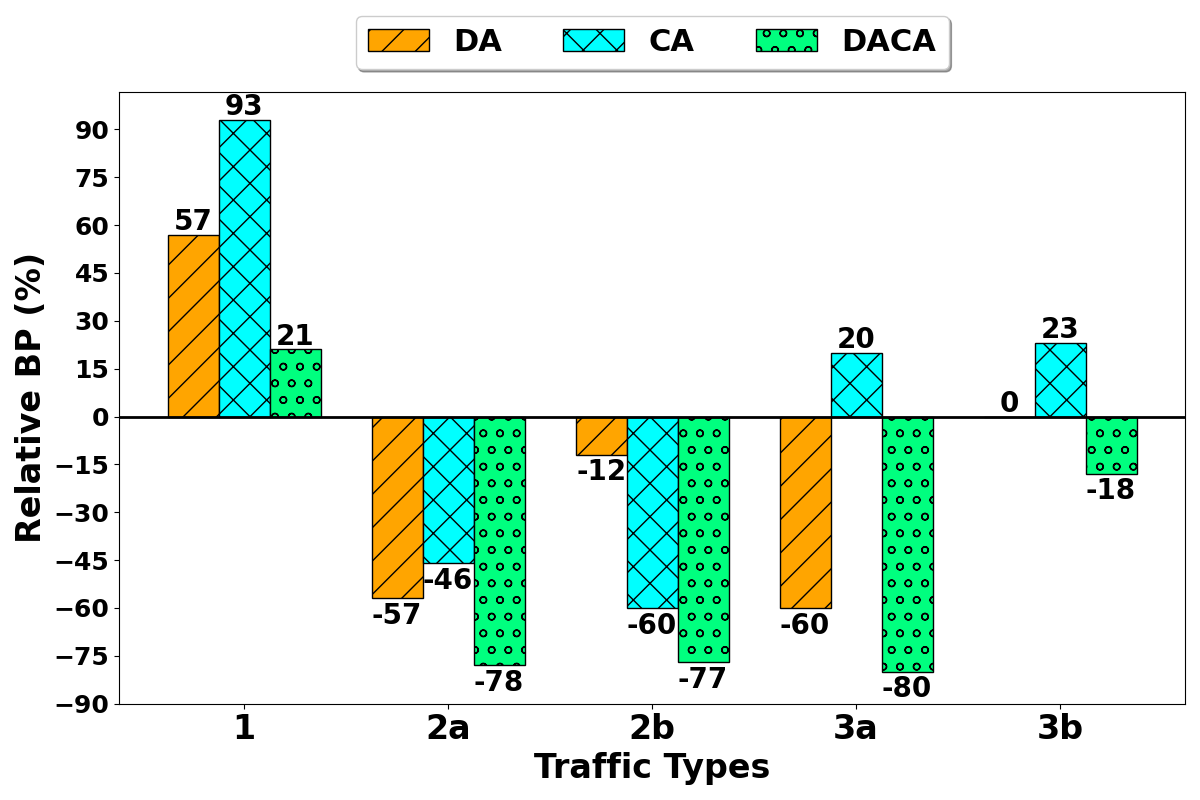}
    % {images_2_day/Relative_BP_new_2_day_with_distribution.png}
    \vspace{-20pt}
    \caption{Relative BP induced by all strategies w.r.t. NDNC per traffic type.}
    \vspace{-5mm}
    \label{fig:Traffic_type_NDNC}
\end{figure}

\vspace{-1mm}
In this section, we assess the impact of each strategy on different traffic types w.r.t.  NDNC. 
In Fig. \ref{fig:Traffic_type_NDNC}, we see that DA reduces the BP of types 2a, 2b, and 3a, by 57\%, 12\%, and 60\%, respectively, and maintains the same BP as NDNC for 3b. Here, DA delays all requests that cannot be provisioned immediately, and hence provisions a higher volume of types 2a, 2b, and 3a, compared to NDNC. Note that traffic from the previous day, particularly type 3b, occupies bandwidth overnight for a significantly longer duration. This results in lack of resources for type 1, leading to its higher BP of 57\%.

% \vspace{-0.1in}
In CA, we observe that the BP for types 2a and 2b is reduced by 46\% and 60\%, respectively. While CA can compress and provision types 2a and 2b, which constitutes 70\% of the total traffic, such provisioning has an adverse effect on other traffic types. Consequently, there is a rise in the BP of traffic types 1, 3a, and 3b by 93\%, 20\%, and 23\%, respectively. Additionally, unlike DA and DACA, type 3b is never delayed, which along with existing traffic from the previous day, contributes to its increased BP and also affects other types. On the other hand, since CA blocks a higher volume of types 2a, 3a, and 3b than DA, it has more resources available for 2b. Thus, the relative BP of type 2b is significantly higher than DA.

In DACA, requests can be provisioned in a compressed form even after their maximum tolerable delay ($\delta$) is exhausted, resulting in 78\% drop in BP for type 2a w.r.t. NDNC. Compressibility also reduces BP for type 2b (non-delayable and compressible) by 77\%. Delayability allows an operator to delay types 2a, 3a, and 3b to ensure spectrum availability, while compressibility reduces the number of channels occupied by types 2a and 2b. The combination of these two factors reduces the relative BP for types 3a and 3b w.r.t. NDNC, while traffic from the previous day leads to a rise in BP of type 1.

\begin{figure}[!t]
    \includegraphics[width=\columnwidth, height=5.0cm]
    % {images/per_hour_3.png}
    {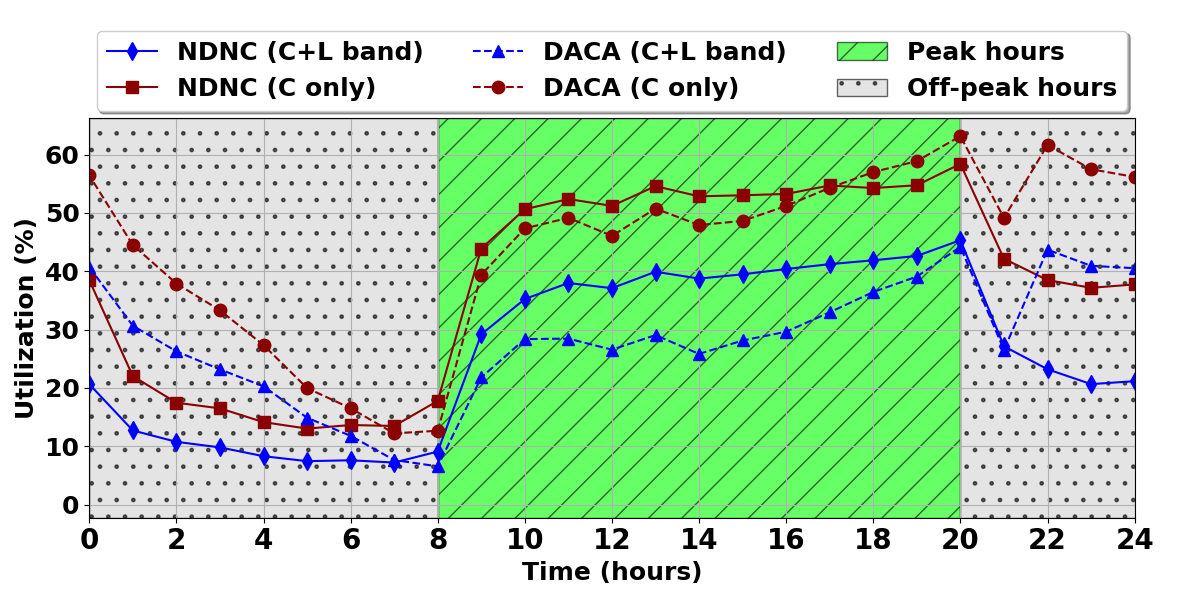}
    \vspace{-23pt}
    \caption{Average resource utilization per hour (NDNC vs. DACA).}
    \label{fig:Utilization_C_C+L}
    \vspace{-5mm}
\end{figure}

To gain a better understanding of resource consumption in the network, Fig. \ref{fig:Utilization_C_C+L} shows the weighted average spectrum utilization per hour for both NDNC and DACA across C-band-only and C+L-band frameworks. We notice that, due to limited bandwidth in C band, utilization is between 50\% and 60\% during peak hours. Note that the uptick observed at 22 hours (10pm) in DACA indicates the bandwidth consumed by type 3b, most of which was delayed until 10pm. However, addition of L band reduces this usage to about 40\% in NDNC. DACA further reduces utilization to about 30\%. While both NDNC and DACA have access to additional bandwidth in C+L, DACA utilizes this bandwidth more effectively by deferring most of type 3b traffic to a later time, allowing other traffic types with much shorter holding times to be served in the network. Overall, these experiments indicate a significant increase in the network's information-carrying capacity.

\vspace{-5pt}
\section{Conclusion}\label{sec:conclusion}
% \vspace{-0.5mm}
% \vspace{-2mm}

We considered a diverse-dynamic traffic scenario in an optical backbone network and proposed a provisioning strategy, called DACA, which exploits delayability and compressibility metrics. Numerical results show that the proposed strategy significantly reduces the blocking probability and hence increases the information-carrying capacity of the network. It also shows that DACA is more effective in C+L framework, offering a promising solution for managing traffic growth in future multi-band networks.

% \vspace{-1mm}
% \section*{Acknowledgment}
% % \vspace{-2.2mm}
% \noindent This work was supported by National Science Foundation (NSF) Grant No. 2226042.

% We appreciate reviewers' feedback which helped us improve this paper.

\bibliographystyle{IEEEtran}

\bibliography{references}

% \bibliography{references_et_al}

\end{document}